\documentclass[prd,showpacs,showkeys,nofootinbib,twocolumn]{revtex4-1}

\usepackage{longtable}

\begin{document}

\title{Equations of motion in gravity theories with nonminimal coupling: A loophole to detect torsion macroscopically?}

\author{Dirk Puetzfeld}
\email{dirk.puetzfeld@zarm.uni-bremen.de}
\homepage{http://puetzfeld.org}
\affiliation{ZARM, University of Bremen, Am Fallturm, 28359 Bremen, Germany} 

\author{Yuri N. Obukhov}
\email{obukhov@ibrae.ac.ru}
\affiliation{Theoretical Physics Laboratory, Nuclear Safety Institute, 
Russian Academy of Sciences, B.Tulskaya 52, 115191 Moscow, Russia}

\date{ \today}

\begin{abstract}
We derive multipolar equations of motion for gravitational theories with general nonminimal coupling in spacetimes admitting torsion. Our very general findings allow for the systematic testing of whole classes of theories by means of extended test bodies. One peculiar feature of certain subclasses of nonminimal theories turns out to be their sensitivity to post-Riemannian spacetime structures even in experiments without microstructured test matter. 
\end{abstract}

\pacs{04.20.Fy; 04.50.Kd; 04.20.Cv}
\keywords{Equations of motion; Conservation laws; Approximation methods}

\maketitle


\section{Introduction}\label{introduction_sec}

In a recent work \cite{Obukhov:Puetzfeld:2013} we derived the conservation laws for the most general class of nonminimally coupled gravity theories. Here we are going to work out the equations of motion for this whole class of theories by using Synge's expansion technique \cite{Synge:1960} in combination with a multipolar framework {\`a} la Dixon \cite{Dixon:1964}. The framework does not only cover the metric case, but it is also general enough to cope with theories which go beyond the usual Riemannian framework \cite{Hehl:1995}. In particular it allows for a generalized discussion of microstructured media. 

The results obtained here extend the ones in \cite{Stoeger:Yasskin:1979,Stoeger:Yasskin:1980,Bertolami:etal:2007,Puetzfeld:Obukhov:2007,Puetzfeld:Obukhov:2008:1,Mohseni:2009,Mohseni:2010,Nojiri:2011,Puetzfeld:Obukhov:2013}. In particular, they offer a new perspective on placing possible observational constraints on new geometric features like torsion.

Our notations and conventions are those of \cite{Hehl:1995}. In particular, the basic geometrical quantities such as the curvature, torsion, etc., are defined as in \cite{Hehl:1995}, and we use the Latin alphabet to label the spacetime coordinate indices. Furthermore, the metric has the signature $(+,-,-,-)$. 

The structure of the paper is as follows: In section \ref{sec_gennonmin} we briefly discuss the class of theories under consideration. In particular we provide the conservation laws, which in turn are crucial for the subsequent derivation of the multipolar equations of motion in section \ref{eom_sec}. Apart from providing the general form of these equations, we study the pole-dipole equations of motion in detail, and thereby find an analogue to the classical Mathisson-Papapetrou \cite{Mathisson:1937,Papapetrou:1951:3} equations for the whole class of nonminimal coupling theories under consideration. Furthermore, we discuss the case of test matter without microstructure and its peculiar type of coupling to post-Riemannian spacetime features. Our final conclusions and an outlook on open problems is given in section \ref{conclusion_sec}. Appendices \ref{conventions_app} and \ref{expansion_app} contain a brief overview of our conventions and some frequently used formulas.

\section{General nonminimal gravity} \label{sec_gennonmin}

In order to be as general as possible, we consider matter with microstructure, namely, with spin. An appropriate gravitational model is then the Poincar\'e gauge theory in which the metric tensor $g_{ij}$ is accompanied by the connection $\Gamma_{ki}{}^j$ that is metric-compatible but not necessarily symmetric. The gravitational field strengths are the Riemann-Cartan curvature and the torsion:
\begin{eqnarray}
R_{kli}{}^j &=& \partial_k\Gamma_{li}{}^j - \partial_l\Gamma_{ki}{}^j + \Gamma_{kn}{}^j \Gamma_{li}{}^n - \Gamma_{ln}{}^j\Gamma_{ki}{}^n,\label{curv}\\
T_{kl}{}^i &=& \Gamma_{kl}{}^i - \Gamma_{lk}{}^i.\label{tors}
\end{eqnarray}

In \cite{Obukhov:Puetzfeld:2013}, we worked out the conservation laws for a general nonminimal gravity model in which the interaction Lagrangian reads
\begin{eqnarray}
L_{\rm int} =  F(g_{ij},R_{kli}{}^j,T_{kl}{}^i) L_{\rm mat}.\label{ansatz_lagrangian_model_3}
\end{eqnarray}
The coupling function $F(g_{ij},R_{kli}{}^j,T_{kl}{}^i)$ depends arbitrarily on its arguments. In technical terms, $F$ is a function of independent scalar invariants constructed in all possible ways from the components of the curvature and torsion tensors.  The matter Lagrangian has the usual form $L_{\rm mat} = L_{\rm mat}(\psi^A, \nabla_i\psi^A, g_{ij})$.

A Lagrange-Noether analysis, see \cite{Obukhov:Puetzfeld:2013}, yields the following conservations laws:
\begin{eqnarray}
F\Sigma_k{}^i &=& Ft_k{}^i + {\stackrel * \nabla}{}_n\left(F\tau^i{}_k{}^n\right), \label{cons1b}\\
{\stackrel * \nabla}{}_i\left(F\Sigma_k{}^i\right) &=& F\Sigma_l{}^i T_{ki}{}^l - F\tau^m{}_n{}^l R_{klm}{}^n\nonumber\\ 
&& -\, L_{\rm mat}\nabla_kF.\label{cons2b}
\end{eqnarray}
Here we made use of the following abbreviations, i.e.\
\begin{equation}
\Sigma_k{}^i = {\frac {\partial {L_{\rm mat}}}{\partial\nabla_i\psi^A}} \,\nabla_k\psi^A - \delta^i_kL_{\rm mat},\label{emcan}
\end{equation}
for the canonical energy-momentum tensor, 
\begin{equation}
\tau^n{}_k{}^i = -\,{\frac {\partial {L_{\rm mat}}}{\partial\nabla_i\psi^A}} \,(\sigma^A{}_B)_k{}^n \psi^B,\label{spin}
\end{equation}
for the canonical spin tensor, and
\begin{equation}\label{emmet}
t_{ij} = {\frac 2{\sqrt{-g}}}\,{\frac {\partial {(\sqrt{-g}L_{\rm mat})}}{\partial g^{ij}}},
\end{equation}
for the metrical energy-momentum tensor. Furthermore, we made use of the so-called modified covariant derivative, which is defined as usual by
\begin{equation}
{\stackrel * \nabla}{}_i = \nabla_i - T_{ki}{}^k.\label{dstar}
\end{equation} 
Lowering the index in (\ref{cons1b}) and antisymmetrizing, we derive the conservation law for the spin
\begin{equation}
F\Sigma_{[ij]} + {\stackrel * \nabla}{}_n\left(F\tau_{[ij]}{}^n\right) = 0.
\end{equation}
This is a generalization of the usual conservation law of the total angular momentum for the case of nonminimal coupling.

\subsection{Purely Riemannian theory}\label{Riemannian_subsec}

Our results contain the Riemannian theory as a special case. Suppose the torsion is absent $T_{ij}{}^k = 0$. Then for usual matter without microstructure (spinless matter with $\tau^m{}_n{}^i = 0$) the canonical and the metrical energy-momentum tensors coincide, $\Sigma_k{}^i = t_k{}^i$. As a result, the conservation law (\ref{cons2b}) reduces to
\begin{equation}
\nabla_it_k{}^i = {\frac 1F}\left(- L_{\rm mat}\delta_k^i -t_k{}^i\right)\nabla_iF.\label{consF}
\end{equation}

\subsection{Further generalization: Matter with intrinsic moments}\label{quadrupole}

Our formalism allows one to consider also the case when matter couples to the gravitational field strengths not just through an $F$-factor in front of the Lagrangian but directly via Pauli-type interaction terms in $L_{\rm mat}$:
\begin{equation}
I^{klm}{}_n(\psi^A, g_{ij})R_{klm}{}^n + J^{kl}{}_n(\psi^A, g_{ij})T_{kl}{}^n.\label{IJ}
\end{equation}
In Maxwell's electrodynamics similar terms describe the interaction of the electromagnetic field to the anomalous magnetic and/or electric dipole moments. For Dirac spinor matter \cite{Obukhov:1998,Hehl:etal:1998}, the Pauli-type quantities $I^{klm}{}_n(\psi^A, g_{ij})$ and $J^{kl}{}_n(\psi^A, g_{ij})$ are interpreted as the (Lorentz and translational, respectively) ``gravitational moments'' that arise from the Gordon decomposition of the dynamical currents.

The on-shell conservation laws are then given by:
\begin{eqnarray}
\Sigma_k{}^i &=& t_k{}^i + {\stackrel * \nabla}{}_n\tau^i{}_k{}^n - 2J^{il}{}_nT_{kl}{}^n + J^{ln}{}_kT_{ln}{}^i \nonumber\\
&& - \,2I^{ilnm}R_{klnm} - 2I^{lnm[i}R_{|lnm|k]},\label{q1a}\\
{\stackrel * \nabla}{}_i\Sigma_k{}^i &=& \Sigma_l{}^i T_{ki}{}^l - \tau^m{}_n{}^l R_{klm}{}^n\nonumber\\ 
&& - \,I^{iln}{}_m\nabla_kR_{iln}{}^m - J^{ln}{}_m\nabla_kT_{ln}{}^m.\label{q2a}
\end{eqnarray}
The skew-symmetric part of (\ref{q1a}) describes the generalized conservation of the angular momentum:
\begin{eqnarray}
{\stackrel * \nabla}{}_n\tau_{[ik]}{}^n &=& -\,\Sigma_{[ik]} + J_{ln[i}T^{ln}{}_{k]} + 2J_{[i}{}^{ln}T_{k]ln}\nonumber\\
&& +\,2I_{[i}{}^{lnm}R_{k]lnm} + 2I^{lnm}{}_{[i}R^{lmn}{}_{k]}.\label{conangmom}
\end{eqnarray}
For the Riemann-Cartan curvature tensor the pairs of indices do not commute, $R_{ijkl} \neq R_{klij}$, and one cannot reduce the two terms in the second line of (\ref{conangmom}). 

However, in the purely Riemannian case of General Relativity, the torsion vanishes and the curvature tensor has more symmetries (in particular, the pairs of indices do commute). Then the system (\ref{q2a}) and (\ref{conangmom}) reduces to the familiar Mathisson-Papapetrou form 
\begin{eqnarray}
\nabla_n\tau_{[ik]}{}^n &=& -\,\Sigma_{[ik]} + 4I_{[i}{}^{lnm}R_{k]lnm},\label{q1b}\\
\nabla_i\Sigma_k{}^i &=&  - \tau^m{}_n{}^l R_{klm}{}^n - I^{ilnm}\nabla_kR_{ilnm}.\label{q2b}
\end{eqnarray}
The symmetric part of equation (\ref{q1b}) describes the relation between the metrical and canonical energy-momentum tensors. When deriving (\ref{q1b}), we took into account that in view of the contraction in (\ref{IJ}), we have the symmetry properties 
\begin{equation}
I^{ijkl} = I^{[ij]kl} = I^{ij[kl]} = I^{klij}.\label{Isym} 
\end{equation}

The form of the system of conservation laws (\ref{q1b})-(\ref{q2b}) is very close to Dixon's equations describing the dynamics of material body with the dipole and quadrupole moments. However, it is important to stress that in contrast to Dixon's {\it integrated} moments of usual structureless matter, $\tau_{[ik]}{}^n$ and $I^{ilnm}$ are the {\it intrinsic} spin and quadrupole moments of matter with microstructure. The above conservation laws can also be viewed as a direct generalization of the ones for spinning particles and polarized media given in \cite{Bailey:Israel:1975}.

It is worthwhile to note that in the Riemann-Cartan spacetime the conservations laws (\ref{q2a}) and (\ref{conangmom}) contain two types of intrinsic quadrupole moments. We identify $I^{ijkl}$ with the {\it rotational} (Lorentz) quadrupole moment, whereas $J^{kl}{}_i$ is naturally interpreted as the {\it translational} quadrupole moment. These quantities are coupled to the corresponding rotational and translational gravitational field strengths, i.e., to the curvature $R_{ijkl}$ and the torsion $T_{kl}{}^i$, respectively.

\section{Equations of motion}\label{eom_sec}

The conservation equations (\ref{cons1b}) and (\ref{cons2b}) form the basis for a general multipolar analysis. In the following we are going to derive the equations of motion for test bodies by utilizing the expansion technique of Synge \cite{Synge:1960}. Since we are now working in a spacetime which allows for more structure, we now also have -- apart from the metric $g_{ab}$ -- the torsion $T_{ab}{}^c$. This leads to an additional degree of freedom regarding the transport operations in the underlying multipolar formalism. We can proceed in two ways: (i) extend Synge's technique to non-Riemannian spacetimes -- thereby switching to a new type of (non-geodesic) reference curve; or (ii) use the standard Riemannian approach and treat torsion as an additional variable. Here we follow the latter strategy.

\subsection{Rewriting conservation laws}\label{Riemannian}

The Riemann-Cartan connection can be decomposed into the Riemannian (Christoffel) connection
\begin{equation}
\widehat{\Gamma}_{ij}{}^k = \left\{{}_{ij}^k\right\} = {\frac 12}g^{kl}\left(\partial_ig_{jl} + \partial_jg_{il} - \partial_lg_{ij}\right),\label{chr}
\end{equation} 
plus the post-Riemannian piece: 
\begin{equation}
{\Gamma}_{ij}{}^k = \widehat{\Gamma}_{ij}{}^k - K_{ij}{}^k.\label{GG}
\end{equation} 
Here the contortion tensor reads
\begin{equation}\label{RCconn}
K_{ij}{}^k = -\,{\frac 12}(T_{ij}{}^k - T_{j}{}^k{}_i + T^k{}_{ij}) = - K_{i}{}^k{}_j .
\end{equation}
We use the hat to denote objects and operators (such as the curvature, covariant derivatives, etc) defined by the Riemannian connection (\ref{chr}).

Using the decomposition (\ref{GG}), we rewrite the conservation laws (\ref{cons1b})-(\ref{cons2b}) as
\begin{eqnarray}
\widehat{\nabla}_n\left(F\tau_{[ik]}{}^n\right) &=& F(K_{ni}{}^l\tau_{[kl]}{}^n - K_{nk}{}^l\tau_{[il]}{}^n)\nonumber\\
&& -\,F\Sigma_{[ik]},\label{cons1c}\\
\widehat{\nabla}_i\left(F\Sigma_k{}^i\right) &=& -\,F\Sigma_l{}^i K_{ki}{}^l - F\tau^m{}_n{}^l R_{klm}{}^n\nonumber\\ 
&& -\, L_{\rm mat}\nabla_kF.\label{cons2c}
\end{eqnarray}
We can develop the usual Riemannian world-function based multipole expansion starting from (\ref{cons1c}) and (\ref{cons2c}). 

Defining auxiliary variables like in \cite{Puetzfeld:Obukhov:2013}, i.e.\ $A(g_{ij},R_{ijk}{}^l,T_{ij}{}^k):=\log F$, $A_i:=\nabla_i A$, $A_{ij}:=\widehat{\nabla}_j \nabla_i A$ etc., we rewrite (\ref{cons1c}) and (\ref{cons2c}) as follows: 
\begin{eqnarray}
\widehat{\nabla}_n \tau_{[ik]}{}^n &=& K_{ni}{}^l \tau_{[kl]}{}^n - K_{nk}{}^l \tau_{[il]}{}^n - \Sigma_{[ik]} \nonumber \\
&&- A_n \tau_{[ik]}{}^n, \label{re_cons1c}\\
\widehat{\nabla}_i \Sigma_k{}^i &=& - \Sigma_l{}^i K_{ki}{}^l - \tau^m{}_n{}^l R_{klm}{}^n - A^i \Xi_{ik} \nonumber \\ 
&&- A_i \Sigma_k{}^i. \label{recons2c}
\end{eqnarray}
Here we introduced the shortcut $\Xi_{ij}:=g_{ij}L_{\rm mat}$.

\subsection{Multipolar approximation}\label{multipolar_approx_subsec}

We will now derive the equations of motion of a test body by utilizing the covariant expansion method of Synge \cite{Synge:1960}. For this we need the following auxiliary formula for the absolute derivative of the integral of an arbitrary bitensor density $\widetilde{B}^{x_1 y_1}=\widetilde{B}^{x_1 y_1}(x,y)$ (the latter is a tensorial function of two spacetime points):
\begin{eqnarray}
{\frac{D}{ds}} \int\limits_{\Sigma(s)} \widetilde{B}^{x_1 y_1} d \Sigma_{x_1} &=& \int\limits_{\Sigma(s)} \widehat{\nabla}_{x_1} \widetilde{B}^{x_1 y_1} w^{x_2} d \Sigma_{x_2} \nonumber \\
&& + \int\limits_{\Sigma(s)} v^{y_2} \widehat{\nabla}_{y_2} \widetilde{B}^{x_1 y_1} d \Sigma_{x_1}.\label{int_aux}
\end{eqnarray}
Here $v^{y_1}:=dx^{y_1}/ds$, $s$ is the proper time, ${\frac{D}{ds}} = v^i\widehat{\nabla}_i$, and the integral is performed over a spatial hypersurface. Note that in our notation the point to which the index of a bitensor belongs can be directly read from the index itself; e.g., $y_{n}$ denotes indices at the point $y$. Furthermore, we will now associate the point $y$ with the world-line of the test body under consideration. Denote 
\begin{eqnarray}
\Phi^{y_1\dots y_ny_0}{}_{x_0} &:=& \sigma^{y_1} \cdots \sigma^{y_n} g^{y_0}{}_{x_0},\label{Phi}\\
\Psi^{y_1\dots y_ny_0y'}{}_{x_0x'} &:=& \sigma^{y_1} \cdots \sigma^{y_n} g^{y_0}{}_{x_0}g^{y'}{}_{x'}.
\label{Psi}
\end{eqnarray}
We start by integrating (\ref{re_cons1c}) and (\ref{recons2c}) using (\ref{int_aux}):
\begin{eqnarray}
&&{\frac{D}{ds}} \int \Psi^{y_1\dots y_ny_0y'}{}_{x_0x'}\widetilde{\tau}^{[x_0 x']x_2} d\Sigma_{x_2} = \nonumber \\
&&\int \Psi^{y_1\dots y_ny_0y'}{}_{x_0x'}\left[ K_{x''x'''}{}^{x_0}\widetilde{\tau}^{[x''' x']x''}\right.\nonumber\\ 
&-& \left. K_{x''x'''}{}^{x'} \widetilde{\tau}^{[x''' x_0]x''} - \widetilde{\Sigma}^{[x_0 x']} - A_{x''} \widetilde{\tau}^{[x_0 x'] x''}\right] w^{x_2} d\Sigma_{x_2}\nonumber\\ 
&+& \int\Psi^{y_1\dots y_ny_0y'}{}_{x_0x';x''}\widetilde{\tau}^{[x_0 x']x''}w^{x_2} d\Sigma_{x_2}\nonumber\\ \label{int_eom_1}
&+& \int v^{y_{n+1}}\Psi^{y_1\dots y_ny_0y'}{}_{x_0x';y_{n+1}}\widetilde{\tau}^{[x_0 x']x_2}d\Sigma_{x_2}, \\
&&{\frac{D}{ds}} \int\Phi^{y_1\dots y_ny_0}{}_{x_0} \widetilde{\Sigma}^{x_0 x_2} d\Sigma_{x_2} = \nonumber \\
&&\int \Phi^{y_1\dots y_ny_0}{}_{x_0} \left[ K^{x_0}{}_{x'x''} \widetilde{\Sigma}^{x' x''} - R^{x_0}{}_{x''' x' x''} \widetilde{\tau}^{x' x'' x'''}\right.\nonumber\\ &-&\left. A_{x'} \left( \widetilde{\Xi}^{x_0x'} + \widetilde{\Sigma}^{x_0x'} \right)\right] w^{x_2} d\Sigma_{x_2}\nonumber\\
&+& \int\Phi^{y_1\dots y_ny_0}{}_{x_0;x'}\widetilde{\Sigma}^{x_0 x'}w^{x_2} d\Sigma_{x_2}\nonumber\\ &+& \int v^{y_{n+1}}\Phi^{y_1\dots y_ny_0}{}_{x_0;y_{n+1}} \widetilde{\Sigma}^{x_0 x_2} d\Sigma_{x_2}.\label{int_eom_2} 
\end{eqnarray}
Here the derivatives are straightforwardly evaluated:
\begin{eqnarray}
&&\Psi^{y_1\dots y_ny_0y'}{}_{x_0x';z} = \sum^{n}_{a=1}\sigma^{y_1}\cdots\sigma^{y_a}{}_z\cdots\sigma^{y_n}g^{y_0}{}_{x_0}g^{y'}{}_{x'}\nonumber\\
&& +\,\sigma^{y_1} \cdots \sigma^{y_n}\left(g^{y_0}{}_{x_0;z}g^{y'}{}_{x'} + g^{y_0}{}_{x_0}g^{y'}{}_{x';z}\right)\label{dPsi},\\
&&\Phi^{y_1\dots y_ny_0}{}_{x_0;z} = \sum^{n}_{a=1}\sigma^{y_1}\cdots\sigma^{y_a}{}_z\cdots\sigma^{y_n}g^{y_0}{}_{x_0}\nonumber\\
&& +\,\sigma^{y_1} \cdots \sigma^{y_n}\,g^{y_0}{}_{x_0;z},\label{dPhi}
\end{eqnarray}
where $z$ stands either for $x$ or for $y$.

We now introduce integrated moments \`{a} la Dixon in \cite{Dixon:1964}, i.e.
\begin{widetext}
\begin{eqnarray}
p^{y_1 \dots y_n y_0}&:=& (-1)^n  \int\limits_{\Sigma(s)}\Phi^{y_1\dots y_ny_0}{}_{x_0} \widetilde{\Sigma}^{x_0 x_1} d \Sigma_{x_1}, \label{p_moments_def} \\
t^{y_2 \dots y_{n+1} y_0 y_1}&:=& (-1)^{n}  \int\limits_{\Sigma(s)} \Psi^{y_2\dots y_{n+1}y_0y_1}{}_{x_0x_1} \widetilde{\Sigma}^{x_0 x_1} w^{x_2} d \Sigma_{x_2}, \label{t_moments_def} \\
\xi^{y_2 \dots y_{n+1} y_0 y_1}&:=& (-1)^{n}  \int\limits_{\Sigma(s)} \Psi^{y_2\dots y_{n+1}y_0y_1}{}_{x_0x_1} \widetilde{\Xi}^{x_0 x_1} w^{x_2} d \Sigma_{x_2}, \label{xi_moments_def} \\
s^{y_2 \dots y_{n+1} y_0 y_1}&:=& (-1)^{n}  \int\limits_{\Sigma(s)} \Psi^{y_2\dots y_{n+1}y_0y_1}{}_{x_0x_1}  \widetilde{\tau}^{[x_0 x_1] x_2} d \Sigma_{x_2}, \label{s_moments_def}\\
q^{y_3 \dots y_{n+2} y_0 y_1 y_2}&:=& (-1)^{n}  \int\limits_{\Sigma(s)} \Psi^{y_3\dots y_{n+2}y_0y_1}{}_{x_0x_1} g^{y_2}{}_{x_2}  \widetilde{\tau}^{[x_0 x_1] x_2} w^{x_3} d \Sigma_{x_3}. \label{q_moments_def}
\end{eqnarray}
Then (\ref{int_eom_1}) and (\ref{int_eom_2}) take the form
\begin{eqnarray}
{\frac{D}{ds}} s^{y_1 \dots y_n y_a y_b} &=& -\,t^{y_1 \dots y_n [y_a y_b]} + q^{(y_1 \dots y_{n-1} |y_a y_b| y_n)} - v^{(y_1}s^{y_2 \dots y_n) y_a y_b} \nonumber\\
&& +\left(v^{y''}s^{y_1 \dots y_{n+1} y' [y_a} + q^{y_1 \dots y_{n+1} y' [y_a |y''|}\right)\widehat{R}{}^{y_b]}{}_{y'y''y_{n+1}} - 2q^{y_1 \dots y_{n+1}[y_a|y'|}K_{y'y_{n+1}}{}^{y_b]}\nonumber\\ 
&& -\,2q^{y_1 \dots y_{n+2} [y_a|y'|}K_{y'y_{n+2}}{}^{y_b]}{}_{;y_{n+1}} - q^{y_1 \dots y_n y_ay_by'}A_{y'} - q^{y_1 \dots y_{n+1} y_ay_by'}A_{y';y_{n+1}}\nonumber\\
&& +\,\sum^{\infty}_{k=2} {\frac{1}{k!}}\left[- q^{y_1 \dots y_{n+k} y_ay_by'}A_{y';y_{n+1}\dots y_{n+k}}
- 2q^{y_1 \dots y_{n+k+1} [y_a|y'|}K_{y'y_{n+k+1}}{}^{y_b]}{}_{;y_{n+1}\dots y_{n+k}} \right.\nonumber\\
&& +\,(-1)^kv^{y'}\beta^{(y_1}{}_{y'y_{n+1}\dots y_{n+k}} s^{y_2 \dots y_n) y_{n+1}\dots y_{n+k}y_a y_b} 
- (-1)^k\alpha^{(y_1}{}_{y'y_{n+1}\dots y_{n+k}} q^{y_2 \dots y_n) y_{n+1}\dots y_{n+k}y_a y_by'}\nonumber\\ 
&&\left. +\,(-1)^k2\left(v^{y'}s^{y_1 \dots y_{n+k+2} [y_a} + q^{y_1 \dots y_{n+k+2}  [y_a |y'|}\right)\gamma^{y_b]}{}_{y_{n+k+2}y'y_{n+1}\dots y_{n+k+1}}\right],\label{int_eom_1_rewritten} \\
{\frac{D}{ds}} p^{y_1 \dots y_{n} y_0} &=& -\,v^{(y_1}p^{y_2 \dots y_{n}) y_0} + t^{(y_1 \dots y_{n-1} |y_0|y_n)} + K^{y_0}{}_{y' y''} t^{y_1 \dots y_{n} y' y''} + K^{y_0}{}_{y' y'';y_{n+1}} t^{y_1 \dots y_{n+1} y' y''}\nonumber\\
&& -\,R^{y_0}{}_{y_{n+1} y' y''} q^{y_1 \dots y_{n} y' y'' y_{n+1}} - R^{y_0}{}_{y_{n+2} y' y'';y_{n+1}} q^{y_1 \dots y_{n+1} y' y'' y_{n+2}} \nonumber\\
&& -\,{\frac 12}\widehat{R}{}^{y_0}{}_{y'y''y_{n+1}}\left(v^{y''}p^{y_1 \dots y_{n+1} y'} + t^{y_1 \dots y_{n+1} y' y''}\right) - A_{y'} \left(\xi^{y_1 \dots y_{n} y'y_0} + t^{y_1 \dots y_{n} y'y_0} \right)\nonumber\\ 
&& -\,A_{y';y''} \left(\xi^{y_1 \dots y_{n} y''y' y_0} + t^{y_1 \dots y_{n} y''y' y_0} \right) \nonumber +\,\sum^{\infty}_{k=2} \frac{1}{k!}\left[ K^{y_0}{}_{y' y'';y_{n+1}\dots y_{n+k}}t^{y_1 \dots y_{n+k} y' y''}\right.\nonumber \\
&& -\,R^{y_0}{}_{y_{n+k+1} y' y'';y_{n+1}\dots y_{n+k}} q^{y_1 \dots y_{n+k} y' y'' y_{n+k+1}}
- A_{y';y_{n+1} \dots y_{n+k}} \left(\xi^{y_1 \dots y_{n+k} y'y_0} + t^{y_1 \dots y_{n+k}y'y_0}\right) \nonumber\\
&& -\,(-1)^k\alpha^{(y_1}{}_{y' y_{n+1} \dots y_{n+k}} t^{y_2 \dots y_n)y_{n+1}\dots y_{n+k}y' y_0} 
+ (-1)^kv^{y'}\beta^{(y_1}{}_{y'y_{n+1}\dots y_{n+k}} p^{y_2 \dots y_n)y_{n+1}\dots y_{n+k} y_0} \nonumber \\
&&\left. -\,(-1)^k\gamma^{y_0}{}_{y'y''y_{n+1}\dots y_{n+k+1}}\left(v^{y''}p^{y_1 \dots y_{n+k+1} y'} + t^{y_1 \dots y_{n+k+1} y'y''} \right)\right].\label{int_eom_2_rewritten} 
\end{eqnarray}
\end{widetext}

\subsection{Vanishing spin current}\label{van_spin_subsec}

For the {\it special case} of vanishing spin current $\tau^{abc}=0$, we infer from (\ref{re_cons1c}) that the canonical energy-momentum tensor is symmetric $\Sigma_{[ij]} = 0$, and that it coincides with the metrical energy-momentum tensor in view of (\ref{cons1b}). Furthermore, we have as a starting point for the derivation of the equations of motion
\begin{eqnarray}
\widehat{\nabla}_i \Sigma^{ki}&=& - K^{k}{}_{il} \Sigma^{li} - A_i \left( \Xi^{ik} + \Sigma^{ki} \right). \label{recons2c_van_spin}
\end{eqnarray}
Due to the antisymmetry of the contortion, the contraction in the first term with the symmetric $t$ moment -- in the case of an absent spin current -- vanishes identically. Hence we are left with structurally the same equation as in \cite{Puetzfeld:Obukhov:2013}, the only\footnote{Note the different sign of $\Xi$ in this paper; this is explained by a different definition of the metrical energy-momentum tensor as compared to \cite{Puetzfeld:Obukhov:2013}.} difference being that here $A(g_{ij},R_{ijk}{}^l,T_{ij}{}^{k})$ is a function of the curvature {\it and} the torsion. 

For the vanishing spin all the corresponding multipole moments (\ref{s_moments_def}) and (\ref{q_moments_def}) vanish, too: $s^{y_2 \dots y_{n+1} y_0 y_1} = 0$ and $q^{y_3 \dots y_{n+1} y_0 y_1y_2} = 0$ for any $n$. In addition, the multipole moments $t^{y_2 \dots y_{n+1} y_0 y_1}$ are symmetric in the last two indices.

\subsubsection{Monopole order ($\tau_{ab}{}^c=0$)}

At the monopole order we have
\begin{eqnarray}
{\frac{D}{ds}} p^{a} &=& - A_{b} \left(\xi^{ab} + t^{ab} \right)\label{mono_eom_1}, \\
t^{ab} &=& p^{a}v^{b}  \label{mono_eom_2}.
\end{eqnarray}
Substituting (\ref{mono_eom_2}) into (\ref{mono_eom_1}), we recover the equation of motion \cite{Puetzfeld:Obukhov:2013}
\begin{equation}
{\frac{D}{ds}} (Fp^{a}) = -\,\xi^{ab}\nabla_bF.\label{mono_eom_3}
\end{equation}
As we see, the nonminimal coupling is manifest in the {\it nongeodetic} motion of the monopole test particle. 

\subsubsection{Pole-dipole order without spin ($\tau_{ab}{}^c=0$)}

At the pole-dipole order we obtain
\begin{eqnarray}
v^{(a} p^{b)c} &=& t^{(ab)c},\label{dipol_eom_1}\\
\frac{D}{ds} p^{ab} &=& t^{ab} - v^{a} p^{b}-\,A_{c} \left(\xi^{abc} + t^{abc} \right),\label{dipol_eom_2} \\
{\frac{D}{ds}} p^{a} &=& -\,{\frac 12} \widehat{R}^{a}{}_{bcd} \left( v^{c} p^{db} + t^{dbc} \right)  -\,A_{b} \left(\xi^{ab} + t^{ab} \right)\nonumber \\ 
&&-\,A_{bc}\left(\xi^{cab} + t^{cab} \right).\label{dipol_eom_3}
\end{eqnarray}
Note that we did not make any simplifying assumptions about the spacetime which still has the general Riemann-Cartan geometric structure with nontrivial torsion. Nevertheless, neither torsion nor contortion contributes to the equations of motion (\ref{dipol_eom_2}) and (\ref{dipol_eom_3}).

\subsection{General pole-dipole equations of motion}\label{dipole_subsec}

Let us consider the general case when the extended body consists of material elements with microstructure, i.e., with spin. In the pole-dipole approximation, the relevant moments\footnote{Note that this counting scheme is compatible with our previous work \cite{Puetzfeld:Obukhov:2007} on multipolar approximations with microstructured matter, in particular it also matches the one employed in \cite{Stoeger:Yasskin:1980}.} are $p^a, p^{ab}, t^{ab}, t^{abc}, \xi^{ab}, \xi^{abc}, s^{ab}, q^{abc}$, and we neglect all higher multipole moments. Then for $n = 1$ and $n = 0$, eq. (\ref{int_eom_1_rewritten}) yields
\begin{eqnarray}
0 &=& -\,t^{a[bc]} + q^{bca} - v^as^{bc},\label{D1}\\
{\frac{D}{ds}} s^{ab} &=& -\,t^{[ab]} - 2q^{c[a|d|}K_{dc}{}^{b]} - q^{abc}A_c,\label{D2} 
\end{eqnarray}
whereas (\ref{int_eom_2_rewritten}) for $n = 2$, $n = 1$, and $n = 0$ yields
\begin{eqnarray}
0 &=& -\,v^{(a}p^{b)c} + t^{(a|c|b)},\label{D3}\\
{\frac{D}{ds}} p^{ab} &=& -\,v^ap^b + t^{ba} + K^b{}_{cd}t^{acd}\nonumber\\ 
&& -\,A_c\,(\xi^{acb} + t^{acb}),\label{D4}\\
{\frac{D}{ds}} p^{a} &=& K^a{}_{cd}t^{cd} + K^a{}_{cd;b}t^{bcd}\nonumber\\
&& -\,R^a{}_{bcd}q^{cdb} - {\frac 12}\widehat{R}{}^a{}_{bcd}(v^cp^{db} + t^{dbc})\nonumber\\
&& -\,A_b\,(\xi^{ba} + t^{ba}) - A_{b;c}\,(\xi^{cba} + t^{cba}).\label{D5}
\end{eqnarray}
Combining (\ref{D1}) with (\ref{D3}), we derive 
\begin{eqnarray}
t^{[a|c|b]} &=& v^cp^{[ab]} + v^{[a}p^{|c|b]} - t^{c[ba]} -  t^{a[bc]} + t^{b[ac]}\label{tabc1}\\
&=& v^cp^{[ab]} + v^{[a}p^{|c|b]} + 2t^{c[ab]} -  3t^{[abc]}.\label{tabc2}
\end{eqnarray}
Furthermore, we can substitute (\ref{D1}) into (\ref{tabc1}) and thus express $t^{[a|c|b]}$ in terms of the $p$-, $q$-, and $s$-moments:
\begin{eqnarray}
t^{[a|c|b]} &=& v^c(p^{[ab]} - s^{ab}) + v^{[a}(p^{|c|b]} + 2s^{b]c})\nonumber\\
&& +\,q^{abc} + 2q^{[a|c|b]}.\label{tabc3}
\end{eqnarray}
Antisymmetrizing (\ref{D4}), we find
\begin{eqnarray}
{\frac{D}{ds}} p^{[ab]} &=& -\,v^{[a}p^{b]} + t^{[ba]} + K^{[b}{}_{cd}t^{a]cd}\nonumber\\ 
&& -\,A_c\,(\xi^{[a|c|b]} + t^{[a|c|b]}).\label{D4a}
\end{eqnarray}
Combining this equation with (\ref{D2}), we eliminate $t^{[ab]}$ and using (\ref{D1}) derive
\begin{eqnarray}
{\frac{D}{ds}}\left(p^{[ab]} - s^{ab}\right) &=& -\,v^{[a}(p^{b]} + K^{b]}{}_{cd}s^{cd})\nonumber\\ 
&& +\,q^{cd[a}K^{b]}{}_{cd} + 2q^{c[a|d|}K_{dc}{}^{b]}\nonumber\\
&& +\,A_c\,(q^{abc} - \xi^{[a|c|b]} - t^{[a|c|b]}).\label{D4b}
\end{eqnarray}
Next, substituting (\ref{D1}), (\ref{D2}), and (\ref{tabc2}) into (\ref{D5}), we obtain after some algebra
\begin{eqnarray}
&{\frac{D}{ds}}\left(p^{a} + K^{a}{}_{cd}s^{cd}\right) = \widehat{R}^a{}_{bcd}(p^{[cd]} - s^{cd})v^b&\nonumber\\ 
& + q^{cdb}[\widehat{R}^a{}_{bcd} - R^a{}_{bcd} + K^a{}_{cd;b} - 2K^a{}_{dn}K_{bc}{}^n - K^a{}_{cd}A_b]\nonumber\\
& -\,A_b(\xi^{ba} + t^{ba}) - A_{b;c}(\xi^{cba} + t^{cba}).&\label{D5a}
\end{eqnarray}
We now introduce the {\it integrated orbital angular momentum} and the {\it integrated spin angular momentum} of an extended body as
\begin{equation}
L^{ab} := 2p^{[ab]},\qquad S^{ab} := -\,2s^{ab},\label{LS}
\end{equation}
respectively. 

Then, after a straightforward but rather lengthy computation, we can recast (\ref{D4b}) and (\ref{D5a}) into the final form
\begin{eqnarray}
{\frac{D}{ds}}{\cal J}^{ab} &=& -\,2v^{[a}{\cal P}^{b]} + 2FQ^{cd[a}T_{cd}{}^{b]} + 4FQ^{[a}{}_{cd}T^{b]cd}\nonumber\\
&& -\,\left(4q^{[a|c|b]} + 2\xi^{[a|c|b]}\right)\nabla_cF,\label{eq_rot}\\
{\frac{D}{ds}}{\cal P}^{a} &=& {\frac 12}\widehat{R}^a{}_{bcd}{\cal J}^{cd}v^b + FQ^{bc}{}_d\widehat{\nabla}{}^a T_{bc}{}^d\nonumber\\
&& -\,2q^{bcd}K_{dc}{}^a\nabla_bF + 2Fq^{acd}\nabla_dA_c\nonumber\\ 
&& -\,\xi^{ba}\nabla_bF - \xi^{cba}\widehat{\nabla}_c\nabla_bF.\label{eq_transl}
\end{eqnarray}
Here we defined the total energy-momentum vector and the total angular momentum tensor by
\begin{eqnarray}\label{Pa}
{\cal P}^a &:=& F\left(p^a - {\frac 12}K^a{}_{cd}S^{cd}\right) + \left(p^{ba} - S^{ab}\right)\nabla_bF,\\
{\cal J}^{ab} &:=& F\left(L^{ab} + S^{ab}\right).\label{Jab}
\end{eqnarray}
In addition, we introduced a redefined moment
\begin{equation}
Q^{bca} := {\frac 12}\left(q^{bca} + q^{bac} - q^{cab}\right).\label{Qabc}
\end{equation}
By construction, $Q^{bc}{}_a = - Q^{cb}{}_a$. In the derivation of (\ref{eq_rot}) and (\ref{eq_transl}) we made use of (\ref{D1}), (\ref{tabc3}) and took into account the geometrical identity 
\begin{equation}\label{RR}
\widehat{R}^a{}_{bcd} - R^a{}_{bcd} \equiv K_{bcd;}{}^a + K^a{}_{cd;b} + 2K_{b[c}{}^nK^a{}_{d]n}.
\end{equation} 
The latter can be proved by substituting the decomposition of the Riemann-Cartan connection (\ref{GG}) into the curvature definition (\ref{curv}). Furthermore, it is helpful to notice that $q^{cd[a}K^{b]}{}_{cd}  + 2q^{c[a|d|}K_{dc}{}^{b]} \equiv Q^{cd[a}T_{cd}{}^{b]} + 2Q^{[a}{}_{cd}T^{b]cd}$ and $q^{cdb}K_{bcd;}{}^a \equiv Q^{bc}{}_d\widehat{\nabla}{}^a T_{bc}{}^d$.

The equations of motion (\ref{eq_rot}) and (\ref{eq_transl}) generalize the results obtained in \cite{Puetzfeld:Obukhov:2013} to the case when extended bodies are built of matter with microstructure and move in a Riemann-Cartan spacetime with nontrivial torsion. 

\subsubsection{Minimal coupling}\label{minimimal_dipole_subsubsec}

When the coupling function is constant, $F = 1$, that is for the {\it minimal coupling} case, we obtain 
\begin{equation}\label{noF}
{\cal P}^a = p^a - {\frac 12}K^a{}_{cd}S^{cd},\qquad {\cal J}^{ab} = L^{ab} + S^{ab},
\end{equation}
and the equations of motion 
\begin{eqnarray}
{\frac{D}{ds}}{\cal J}^{ab} &=& -\,2v^{[a}{\cal P}^{b]} + 2Q^{cd[a}T_{cd}{}^{b]} + 4Q^{[a}{}_{cd}T^{b]cd},\label{eq_rot_noF}\\
{\frac{D}{ds}}{\cal P}^{a} &=& {\frac 12}\widehat{R}^a{}_{bcd}{\cal J}^{cd}v^b + Q^{bc}{}_d\widehat{\nabla}{}^a T_{bc}{}^d.\label{eq_transl_noF}
\end{eqnarray}
Comparing these equations to the conservation laws (\ref{q2a}) and (\ref{conangmom}), it is remarkable that the redefined dipole spin moment (\ref{Qabc}) actually took over the role of the translational quadrupole moment. That is, up to a factor $(-2)$, conventionally introduced in (\ref{LS}), we can identify $Q^{bc}{}_a$ with $J^{bc}{}_a$. This interesting feature was not reported before.

\subsubsection{Nonminimal coupling: a loophole to detect torsion?}\label{nonminimimal_dipole_subsubsec}

It is satisfying to see that the structure of the equations of motion (\ref{eq_rot_noF})-(\ref{eq_transl_noF}) is in agreement with the earlier results of Yasskin and Stoeger \cite{Stoeger:Yasskin:1980}. Therefore, we confirm once again that spacetime torsion couples only to the integrated spin $S^{ab}$, which arises from the intrinsic spin of matter, and the higher moment $q^{abc}$. Hence, usual matter without microstructure cannot detect torsion and, in particular, experiments with macroscopically rotating bodies such as gyroscopes in the Gravity Probe B mission do not place any limits on torsion \cite{Hehl:Obukhov:Puetzfeld:2013}. 

However, this conclusion is apparently violated for the {\it nonminimal coupling} case. As we see from (\ref{eq_rot}) and (\ref{eq_transl}), test bodies of structureless matter could be affected by torsion via the derivatives of the coupling function $F(g_{ij},R_{kli}{}^j,T_{kl}{}^i)$. On the other hand, this possibility is qualitatively different from the ad hoc assumption that structureless particles move along auto-parallel curves in the Riemann-Cartan spacetime made in \cite{Kleinert:1998,Mao:2009,March:2011a,March:2011b}; see the critical assessment in \cite{Hehl:Obukhov:Puetzfeld:2013}. The trajectory of a monopole particle, described by (\ref{mono_eom_3}), is neither geodesic nor auto-parallel. The same is true for the dipole case when the nonminimal coupling force is combined with the Mathisson-Papapetrou force.

\section{Conclusion}\label{conclusion_sec}

We have obtained equations of motion for material bodies with microstructure, thus generalizing the previous works \cite{Bailey:Israel:1975,Stoeger:Yasskin:1979,Stoeger:Yasskin:1980,Puetzfeld:Obukhov:2007,Puetzfeld:Obukhov:2008:1} to the general framework with nonminimal coupling. The master equations (\ref{int_eom_1_rewritten}) and (\ref{int_eom_2_rewritten}) describe the dynamics of an extended body up to an arbitrary multipole order. It turns out that, despite a rather complicated general structure of the equations of motion, most of the terms in (\ref{int_eom_1_rewritten}) and (\ref{int_eom_2_rewritten}) show up only at the quadrupole order or higher orders. 

In the special case of minimal coupling (which is recovered when $F=1$), our results can be viewed as the covariant generalization of the ones in \cite{Stoeger:Yasskin:1979,Stoeger:Yasskin:1980}, as well as the parts concerning Poincar\'{e} gauge theory of \cite{Puetzfeld:Obukhov:2007}. 

A somewhat surprising result in the present nonminimal context with torsion, is the -- indirect -- appearance of the torsion through the coupling function $F$ even in the lowest order equations of motion for matter without intrinsic spin -- see eqs.\ (\ref{mono_eom_1})--(\ref{mono_eom_2}). This clearly is a distinctive feature of theories which exhibit nonminimal coupling, which sets them apart from other gauge theoretical approaches to gravity. As we have shown in \cite{Stoeger:Yasskin:1980,Puetzfeld:Obukhov:2007,Puetzfeld:Obukhov:2008:1}, and as it is also discussed at length in the recent review \cite{Hehl:Obukhov:Puetzfeld:2013}, in the minimally coupled case only microstructured matter couples to the post-Riemannian spacetime features -- in particular, in the minimally coupled case one needs matter with intrinsic spin to detect the possible torsion of spacetime. As we have shown in the current work, this is no longer the case in the nonminimally coupled context. In other words, supposing that one can come up with a sensible background model for spacetime including torsion, it could be somewhat constrained through standard test bodies -- i.e.\ made from regular matter -- through the derived equations of motion, in particular through (\ref{mono_eom_1})--(\ref{mono_eom_2}) in the monopolar case. 

Despite the progress made here, we would also like to point out some open questions and directions for future investigations. (i) In a post-Riemannian context, there is naturally more freedom regarding the possible geometry of spacetime. This additional freedom could also be used for an extension and modification of the multipolar framework itself in general spacetimes encompassing, besides the curvature, also new quantities like torsion. In particular, one could carry out the derivations in the present work with a modified world-function formalism, i.e.\ one which is no longer based on the geodesic structure of the spacetime -- see also \cite{Goldthorpe:1980,Nieh:Yan:1982,Barth:1987} for some generalizations in this direction. While such a modification remains a possibility, which is somewhat linked to the discussion of which types of curves are ``natural'' in specific spacetimes, one should also be clear that one would loose comparability with almost all of the previous works on equations of motion. (ii) Another generalization concerns the generalization to the metric-affine case, i.e. including, apart from the torsion, also the nonmetricity of spacetime. The results in this paper already hint into this direction. In general non-Riemannian spacetimes, one can expect a direct coupling term, not only through the function $F$, on the level of the equations of motion. This will eventually lead to more ``fine grained'' possible tests of post-Riemannian geometric structures.

\section*{Acknowledgements}
This work was supported by the Deutsche Forschungsgemeinschaft (DFG) through the grant LA-905/8-1/2 (D.P.). 

\appendix

\section{Conventions \& Symbols}\label{conventions_app}

\begin{table}
\caption{\label{tab_symbols}Directory of symbols.}
\begin{ruledtabular}
\begin{tabular}{ll}
Symbol & Explanation\\
\hline
&\\
\hline
\multicolumn{2}{l}{{Geometrical quantities}}\\
\hline
$g_{a b}$ & Metric\\
$\sqrt{-g}$ & Determinant of the metric \\
$\delta^a_b$ & Kronecker symbol \\
$x^{a}$, $s$ & Coordinates, proper time \\
$\Gamma_{a b}{}^c$ & Connection \\
$K_{a b}{}^c$ & Contortion \\
$T_{a b}{}^c$ & Torsion \\
$R_{a b c}{}^d$& Curvature \\
$\sigma$ & World-function\\
$g^{y_0}{}_{x_0}$ & Parallel propagator\\
&\\
\hline
\multicolumn{2}{l}{{Matter quantities}}\\
\hline
$\psi^A$ & General matter field \\
$\Sigma_a{}^b$ & Canonical energy-momentum \\
$t_a{}^b$ & Metrical energy-momentum \\
$\tau^a{}_b{}^c$ & Canonical spin \\
$I^{abcd}, J^{abc}$ & Pauli-type moments \\
$v^a$ & Velocity \\
$\mathcal{P}^a$ & Generalized momentum \\
$S^{ab}$ & Spin angular momentum \\
$L^{ab}$ & Orbital angular momentum \\
$\mathcal{J}^{ab}$ & Total angular momentum \\
$L$ & Lagrangian \\ 
$p^{y_1 \cdots y_n y_0}$, $t^{y_2 \cdots y_{n+1} y_0 y_1}$,  & Integrated moments\\
$\xi^{y_2 \cdots y_{n+1} y_0 y_1}$, $s^{y_2 \cdots y_{n+1} y_0 y_1}$, &\\
$q^{y_3 \cdots y_{n+2} y_0 y_1 y_2}$ & \\
&\\
\hline
\multicolumn{2}{l}{{Auxiliary quantities}}\\
\hline
${\stackrel * \nabla}{}_a$ & Modified cov. derivative \\
$F$, $A$ & Coupling function\\
$\alpha^{y_0}{}_{y_1 \dots y_n}$, $\beta^{y_0}{}_{y_1 \dots y_n}$, $\gamma^{y_0}{}_{y_1 \dots y_n}$& Expansion coefficients\\
&\\
\hline
\multicolumn{2}{l}{{Operators}}\\
\hline
$\partial_i$, $\nabla_i$ & (Partial, covariant) derivative \\ 
$\frac{D}{ds} = $``$\dot{\phantom{a}}$'' & Total derivative \\
``$[ \dots ]$''& Coincidence limit\\
``$\widehat{\phantom{AA}}$'' & Riemannian quantity
\end{tabular}
\end{ruledtabular}
\end{table}
 
In the following we summarize our conventions, and collect some frequently used formulas. A directory of symbols used throughout the text can be found in table \ref{tab_symbols}.

For an arbitrary $k$-tensor $T_{a_1 \dots a_k}$, the symmetrization and antisymmetrization are defined by
\begin{eqnarray}
T_{(a_1\dots a_k)} &:=& {\frac 1{k!}}\sum_{I=1}^{k!}T_{\pi_I\!\{a_1\dots a_k\}},\label{S}\\
T_{[a_1\dots a_k]} &:=& {\frac 1{k!}}\sum_{I=1}^{k!}(-1)^{|\pi_I|}T_{\pi_I\!\{a_1\dots a_k\}},\label{A}
\end{eqnarray}
where the sum is taken over all possible permutations (symbolically denoted by $\pi_I\!\{a_1\dots a_k\}$) of its $k$ indices. As is well-known, the number of such permutations is equal to $k!$. The sign factor depends on whether a permutation is even ($|\pi| = 0$) or odd ($|\pi| = 1$). The number of independent components of the totally symmetric tensor $T_{(a_1\dots a_k)}$ of rank $k$ in $n$ dimensions is equal to the binomial coefficient ${{n-1+k}\choose{k}} = (n-1+k)!/[k!(n-1)!]$, whereas the number of independent components of the totally antisymmetric tensor $T_{[a_1\dots a_k]}$ of rank $k$ in $n$ dimensions is equal to the binomial coefficient ${{n}\choose{k}} = n!/[k!(n-k)!]$. For example, for a second rank tensor $T_{ab}$ the symmetrization yields a tensor $T_{(ab)} = {\frac 12}(T_{ab} + T_{ba})$ with 10 independent components, and the antisymmetrization yields another tensor $T_{[ab]} = {\frac 12}(T_{ab} - T_{ba})$ with 6 independent components.
 
The covariant derivative defined by the Riemannian connection (\ref{chr}) is conventionally denoted by the nabla or by the semicolon: $\widehat{\nabla}_a =$ ``$ {}_{;a}$''.

Our conventions for the Riemann curvature are as follows:
\begin{eqnarray}
&& 2 A^{c_1 \dots c_k}{}_{d_1 \dots d_l ; [ba] } \equiv 2 \widehat{\nabla}_{[a} \widehat{\nabla}_{b]} A^{c_1 \dots c_k}{}_{d_1 \dots d_l} \nonumber \\
& = & \sum^{k}_{i=1} \widehat{R}_{abe}{}^{c_i} A^{c_1 \dots e \dots c_k}{}_{d_1 \dots d_l} \nonumber \\
&& - \sum^{l}_{j=1} \widehat{R}_{abd_j}{}^{e} A^{c_1 \dots c_k}{}_{d_1 \dots e \dots d_l}. \label{curvature_def}
\end{eqnarray}
The Ricci tensor is introduced by $\widehat{R}_{ij} := \widehat{R}_{kij}{}^k$, and the curvature scalar is $\widehat{R} := g^{ij}\widehat{R}_{ij}$. The signature of the spacetime metric is assumed to be $(+1,-1,-1,-1)$.

In the following, we summarize some of the frequently used formulas in the context of the bitensor formalism (in particular for the world-function $\sigma(x,y)$), see, e.g., \cite{Synge:1960,DeWitt:Brehme:1960,Poisson:etal:2011} for the corresponding derivations. Note that our curvature conventions differ from those in \cite{Synge:1960,Poisson:etal:2011}. Indices attached to the world-function always denote covariant derivatives, at the given point, i.e.\ $\sigma_y:= \nabla_y \sigma$, hence we do not make explicit use of the semicolon in case of the world-function. We start by stating, without proof, the following useful rule for a bitensor $B$ with arbitrary indices at different points (here just denoted by dots):
\begin{eqnarray}
\left[B_{\dots} \right]_{;y} = \left[B_{\dots ; y} \right] + \left[B_{\dots ; x} \right]. \label{synges_rule}
\end{eqnarray}
Here a coincidence limit of a bitensor $B_{\dots}(x,y)$ is a tensor 
\begin{eqnarray}
\left[B_{\dots} \right] = \lim\limits_{x\rightarrow y}\,B_{\dots}(x,y),\label{coin}
\end{eqnarray}
determined at $y$. Furthermore, we collect the following useful identities: 
\begin{eqnarray}
&&\sigma_{y_0 y_1 x_0 y_2 x_1} = \sigma_{y_0 y_1 y_2 x_0 x_1} = \sigma_{x_0 x_1 y_0 y_1 y_2 }, \label{rule_1} \\
&&g^{x_1 x_2} \sigma_{x_1} \sigma_{x_2} = 2 \sigma = g^{y_1 y_2} \sigma_{y_1} \sigma_{y_2}, \label{rule_2}\\
&&\left[ \sigma \right]=0, \quad  \left[ \sigma_x \right] = \left[ \sigma_y \right]  = 0, \label{rule_3} \\
&& \left[ \sigma_{x_1 x_2} \right] =  \left[ \sigma_{y_1 y_2} \right] = g_{y_1 y_2}, \label{rule_4}\\ 
&& \left[ \sigma_{x_1 y_2} \right] =  \left[ \sigma_{y_1 x_2} \right] = - g_{y_1 y_2}, \label{rule_5}\\ 
&& \left[ \sigma_{x_1 x_2 x_3} \right] = \left[ \sigma_{x_1 x_2 y_3} \right] = \left[ \sigma_{x_1 y_2 y_3} \right] = \left[ \sigma_{y_1 y_2 y_3} \right] = 0, \nonumber \\ \label{rule_6}\\
&&\left[g^{x_0}{}_{y_1} \right] = \delta^{y_0}{}_{y_1}, \quad \left[g^{x_0}{}_{y_1 ; x_2} \right] = \left[g^{x_0}{}_{y_1 ; y_2} \right] = 0, \label{rule_7} \\
&& \left[g^{x_0}{}_{y_1 ; x_2 x_3} \right] = \frac{1}{2} \widehat{R}{}^{y_0}{}_{y_1 y_2 y_3}. \label{rule_8}
\end{eqnarray}

\section{Covariant expansions}\label{expansion_app}

Here we briefly summarize the covariant expansions of the second derivative of the world-function, and the derivative of the parallel propagator:
\begin{eqnarray}
\sigma^{y_0}{}_{x_1} &=& g^{y'}{}_{x_1}\biggl( -\,\delta^{y_0}{}_{y'}\nonumber\\
&& +\,\sum\limits_{k=2}^\infty\,{\frac {1}{k!}}\,\alpha^{y_0}{}_{y'y_2\!\dots \!y_{k+1}}\sigma^{y_2}\cdots\sigma^{y_{k+1}}\biggr)\!,\label{app_expansion_1}\\
\sigma^{y_0}{}_{y_1} &=& \delta^{y_0}{}_{y_1} \nonumber\\
&& -\,\sum\limits_{k=2}^\infty\,{\frac {1}{k!}}\,\beta^{y_0}{}_{y_1y_2\dots y_{k+1}} \sigma^{y_2}\!\cdots\!\sigma^{y_{k+1}}, \label{app_expansion_2} \\
g^{y_0}{}_{x_1 ; x_2} &=& g^{y'}{\!}_{x_1} g^{y''}{\!}_{x_2}\biggl({\frac 12} 
\widehat{R}{}^{y_0}{}_{y'y''y_3}\sigma^{y_3}\nonumber\\ 
&&\!+\!\sum\limits_{k=2}^\infty\,{\frac {1}{k!}}\,\gamma^{y_0}{}_{y'y''y_3\dots y_{k+2}}\sigma^{y_3}\!\cdots\!\sigma^{y_{k+2}}\!\biggr)\!,\label{app_expansion_3} \\
g^{y_0}{}_{x_1 ; y_2} &=& g^{y'}{\!}_{x_1} \biggl({\frac 12} \widehat{R}{}^{y_0}{}_{y'y_2y_3}\sigma^{y_3}\nonumber\\ 
&&\!+\!\sum\limits_{k=2}^\infty\,{\frac {1}{k!}}\,\gamma^{y_0}{}_{y'y_2y_3\dots y_{k+2}}\sigma^{y_3}\!\cdots\!\sigma^{y_{k+2}}\!\biggr).\label{app_expansion_4}
\end{eqnarray}
The coefficients $\alpha, \beta, \gamma$ in these expansions are polynomials constructed from the Riemann curvature tensor and its covariant derivatives. The first coefficients read as follows:
\begin{eqnarray}
\alpha^{y_0}{}_{y_1y_2y_3} &=& - \frac{1}{3} \widehat{R}{}^{y_0}{}_{(y_2y_3)y_1},\label{a1}\\
\beta^{y_0}{}_{y_1y_2y_3} &=& \frac{2}{3}\widehat{R}{}^{y_0}{}_{(y_2y_3)y_1},\label{be1}\\
\alpha^{y_0}{}_{y_1y_2y_3y_4} &=& - \frac{1}{2} \widehat{\nabla}_{(y_2}\widehat{R}{}^{y_0}{}_{y_3y_4)y_1},\label{al2}\\
\beta^{y_0}{}_{y_1y_2y_3y_4} &=& \frac{1}{2} \widehat{\nabla}_{(y_2} \widehat{R}{}^{y_0}{}_{y_3y_4)y_1},\label{be2}\\
\nonumber\\
\gamma^{y_0}{}_{y_1y_2y_3y_4}&=& \frac{1}{3} \widehat{\nabla}_{(y_3} \widehat{R}{}^{y_0}{}_{|y_1|y_4)y_2}.\label{ga}
\end{eqnarray}
In addition, we also need the covariant expansion of a usual vector:
\begin{eqnarray} 
A_x = g^{y_0}{}_x\,\sum\limits_{k=0}^\infty\,{\frac {(-1)^k}{k!}} \, A_{y_0;y_1\dots y_k}\,\sigma^{y_1}\cdots\sigma^{y_k}.\label{Ax}
\end{eqnarray}

\bibliographystyle{unsrtnat}
\bibliography{gennonmineom}

\end{document}